\documentclass[dvips]{actaprep}
\usepackage{supertabular,lscape,epsfig}
\usepackage{amssymb}
\usepackage{amsmath}
\DeclareSymbolFont{ppa}{OT1}{ppl}{m}{it}
\DeclareMathSymbol{\vv}{\mathalpha}{ppa}{'166}

\newfont{\hb}{rphvb at 10pt}%bezszeryfowe pó³grube
\newfont{\hbo}{rphvbo at 10pt}%bezszeryfowe pó³grube kursywa
\newfont{\bitt}{rptmbi at 12pt}%pó³gruba kursywa (tytu³ artyku³u)
\newfont{\bits}{rptmbi at 11pt}%pó³gruba kursywa (tytu³y rozdzia³ów)

\renewcommand{\FigCap}[1]{\par\noindent Fig.\  % 
  \refstepcounter{figure}\thefigure. #1\par}

\SetPages{295}{304}

\SetVol{60}{2010}

\begin{document}

%Zwarte naglowki, jeden wiersz, appendix
\newcommand{\TabApp}[2]{\begin{center}\parbox[t]{#1}{\centerline{
  {\bf Appendix}}
  \vskip2mm
  \centerline{\small {\spaceskip 2pt plus 1pt minus 1pt T a b l e}
  \refstepcounter{table}\thetable}
  \vskip2mm
  \centerline{\footnotesize #2}}
  \vskip3mm
\end{center}}

%Zwarte naglowki, jeden wiersz
\newcommand{\TabCapp}[2]{\begin{center}\parbox[t]{#1}{\centerline{
  \small {\spaceskip 2pt plus 1pt minus 1pt T a b l e}
  \refstepcounter{table}\thetable}
  \vskip2mm
  \centerline{\footnotesize #2}}
  \vskip3mm
\end{center}}

%Zwarte naglowki, dwa wiersze
\newcommand{\TTabCap}[3]{\begin{center}\parbox[t]{#1}{\centerline{
  \small {\spaceskip 2pt plus 1pt minus 1pt T a b l e}
  \refstepcounter{table}\thetable}
  \vskip2mm
  \centerline{\footnotesize #2}
  \centerline{\footnotesize #3}}
  \vskip1mm
\end{center}}

%Zwarte naglowki, jeden wiersz, appendix
\newcommand{\MakeTableApp}[4]{\begin{table}[p]\TabApp{#2}{#3}
  \begin{center} \TableFont \begin{tabular}{#1} #4 
  \end{tabular}\end{center}\end{table}}

%Zwarte naglowki, jeden wiersz
\newcommand{\MakeTableSepp}[4]{\begin{table}[p]\TabCapp{#2}{#3}
  \begin{center} \TableFont \begin{tabular}{#1} #4 
  \end{tabular}\end{center}\end{table}}

%Zwarte naglowki, jeden wiersz
\newcommand{\MakeTableee}[4]{\begin{table}[htb]\TabCapp{#2}{#3}
  \begin{center} \TableFont \begin{tabular}{#1} #4
  \end{tabular}\end{center}\end{table}}

%Zwarte naglowki, dwa wiersze
\newcommand{\MakeTablee}[5]{\begin{table}[htb]\TTabCap{#2}{#3}{#4}
  \begin{center} \TableFont \begin{tabular}{#1} #5 
  \end{tabular}\end{center}\end{table}}

%FWHM, PSF - proste, MgII, H$\alpha$
%rms, rhs, sd - kursywa
%{\sc DAOPhot}
%{\sf files}
%Galactic wszystko (bulge, center, plane...)
%Cepheids
%type~ Cepheids, Population~II Cepheids
\newfont{\bb}{ptmbi8t at 12pt}
\newfont{\bbb}{cmbxti10}
\newfont{\bbbb}{cmbxti10 at 9pt}
\newcommand{\uprule}{\rule{0pt}{2.5ex}}
\newcommand{\douprule}{\rule[-2ex]{0pt}{4.5ex}}
\newcommand{\dorule}{\rule[-2ex]{0pt}{2ex}}
\begin{Titlepage}
\Title{The Optical Gravitational Lensing Experiment.\\
OGLE-III Photometric Maps of the Galactic Disk Fields.\footnote{Based
on observations obtained with the 1.3~m Warsaw telescope at the Las
Campanas Observatory of the Carnegie Institution of Washington.}}
\Author{
M.\,K.~~S~z~y~m~a~ñ~s~k~i$^1$,~~
A.~~U~d~a~l~s~k~i$^1$,~~ 
I.~~S~o~s~z~y~\'n~s~k~i$^1$,~~
M.~~K~u~b~i~a~k$^1$,\\
G.~~P~i~e~t~r~z~y~\'n~s~k~i$^{1,2}$,~~ 
R.~~P~o~l~e~s~k~i$^1$,~~
£.~~W~y~r~z~y~k~o~w~s~k~i$^3$~~and~~
K.~~U~l~a~c~z~y~k$^1$}
{$^1$Warsaw University Observatory, Al.~Ujazdowskie~4, 00-478~Warszawa, Poland\\
e-mail: (msz,udalski,soszynsk,mk,pietrzyn,rpoleski,kulaczyk)@astrouw.edu.pl\\
$^2$ Universidad de Concepci{\'o}n, Departamento de Astronomia,
Casilla 160--C, Concepci{\'o}n, Chile\\
$^3$ Institute of Astronomy, University of Cambridge, Madingley Road,
Cambridge CB3~0HA,~UK\\
e-mail: wyrzykow@ast.cam.ac.uk}
\Received{December 22, 2010}
\end{Titlepage}

\Abstract{We present OGLE-III Photometric Maps of the Galactic disk
fields observed during the OGLE-III campaigns for low luminosity transiting
objects that led to the discovery of the first transitng exoplanets.

The maps contain precise, calibrated {\it VI} photometry of about 9 million
stars from 21 OGLE-III fields in the Galactic disk observed in the years
2002--2009 and covering more than 7 square degrees in the sky. Precise
astrometry of these objects is also provided.

We discuss quality of the data and present a few color--magnitude diagrams
of the observed fields.

All photometric data are available to the astronomical community from the
OGLE Internet archive.}{Galaxy: disk -- Surveys -- Catalogs -- Techniques:
photometric}

\Section{Introduction}
Photometry of millions of stars collected during the regular long term
monitoring of the targets of the Optical Gravitational Lensing Experiment
(OGLE) is a unique observational material that can be used for many
astrophysical applications. Since the second phase, OGLE-II, the OGLE group
has regularly released the OGLE Photometric Maps of Dense Stellar Regions
containing calibrated {\it BVI} or {\it VI} photometry and precise
astrometry of millions of stars from the observed fields. They included
astrophysically important objects like the Large and Small Magellanic
Clouds and the Galactic Center.

The most recent version of the OGLE Maps comes from the third phase of the
OGLE project -- OGLE-III. OGLE-III observations covered an area of the
sky larger by an order of magnitude
as compared to the original OGLE-II maps, and contain
photometry of about ten times more stars. So far
OGLE-III Maps of the Large and Small Magellanic Clouds were released
(Udalski \etal 2008ab).

OGLE-III maps have already been widely used by astronomers to many projects
(\eg Subramanian and Subramaniam 2010, Szczygie³ \etal 2010). They are also
widely used as a huge set of secondary photometric standards for
calibrating photometry.

The OGLE-III target list included a set of fields from the Galactic
disk. These fields located in dense stellar regions at low Galactic
lattitude and longitudes between $280\arcd$ and $310\arcd$ were
extensively monitored with high cadence (order of 15 minutes) for low
luminosity object and planetary transits leading to the discovery of the
first transiting exoplanets (Udalski \etal 2002, Bouchy \etal 2004).

Precise photometry of the Galactic disk fields can be a very useful tool
for studying the Galactic structure. Large area of the sky around the
Galactic equator has been currently monitored by the OGLE-IV survey in the
optical, {\it VI}, domain, as well as in the near infrared by the VVV
project conducted on the VISTA telescope at ESO Paranal Observatory,
Chile. OGLE-III photometry of selected Galactic disk fields can be then a
good anchoring point for these larger scale surveys of the Galactic disk.

This paper is the next in the OGLE-III Map series. We present here OGLE-III
photometric maps of the Galactic disk fields covering about 7 square
degrees in the sky and containing photometry and astrometry of about 9
million stars. The maps are available to the astrophysical community from
the OGLE Internet archive.
\vspace*{-7pt}
\Section{Observational Data}
\vspace*{-5pt}
The photometric data presented in this paper were collected during the
OGLE-III phase between February 2002 and May 2009 with the 1.3-m Warsaw
Telescope at Las Campanas Observatory, Chile, operated by the Carnegie
Institution of Washington. The telescope was equipped with the eight chip
mosaic camera (Udalski 2003) covering approximately $35\times35$~arcmin in
the sky with the scale of 0.26 arcsec/pixel.

Observations were carried out in {\it V}- and {\it I}-band filters closely
resembling the standard bands. One should be however aware, that the OGLE
glass {\it I}-band filter approximates well the standard one for
$V-I<3$~mag colors. For very red objects the transformation to the standard
band is less precise. The vast majority of observations were obtained
through the {\it I}-band filter. Typically up to $\sim2700$ images for each
field were collected in this band and just a few in the {\it V}-band. The
exposure time was 180~s or 120~s for the {\it I}-band and 240~s for the
{\it V}-band.

Observations were conducted only in good seeing (less than 1\zdot\arcs8)
and transparency conditions. The median seeing of the {\it I}-band images
is equal to $1\zdot\arcs2$.

The Galactic disk fields observed during OGLE-III phase as well as the
equatorial and galactic coordinates of their centers and number of stars
detected in the {\it I}-band are listed in Table~1. The area of the sky
covered by OGLE-III observations of these fields exceeds 7 square
degrees.

\MakeTableee{ccccrr}{12.5cm}{OGLE-III Galactic Disk Fields}
{\hline
\noalign{\vskip3pt}
Field & RA       &   DEC   & $l_{II}$ & \multicolumn{1}{c}{$b_{II}$} & $N_{\rm Stars}$ \\
    & (2000)   &  (2000) & & & \\
\noalign{\vskip3pt}
\hline
\noalign{\vskip3pt}
CAR100  & 11\uph07\upm00\ups & $-61$\arcd06\arcm30\arcs & 290{\zdot\arcd}6544 &  $-0${\zdot\arcd}7510  & 382528 \\
CAR104  & 10\uph57\upm30\ups & $-61$\arcd40\arcm00\arcs & 289{\zdot\arcd}8439 &  $-1${\zdot\arcd}7249  & 475893 \\
CAR105  & 10\uph52\upm20\ups & $-61$\arcd40\arcm00\arcs & 289{\zdot\arcd}2911 &  $-1${\zdot\arcd}9906  & 459781 \\
CAR106  & 11\uph03\upm00\ups & $-61$\arcd50\arcm00\arcs & 290{\zdot\arcd}5054 &  $-1${\zdot\arcd}6063  & 401528 \\
CAR107  & 10\uph47\upm15\ups & $-62$\arcd00\arcm25\arcs & 288{\zdot\arcd}9089 &  $-2${\zdot\arcd}5647  & 318932 \\
CAR108  & 10\uph47\upm15\ups & $-61$\arcd24\arcm35\arcs & 288{\zdot\arcd}6343 &  $-2${\zdot\arcd}0343  & 379610 \\
CAR109  & 10\uph42\upm10\ups & $-62$\arcd10\arcm25\arcs & 288{\zdot\arcd}4607 &  $-2${\zdot\arcd}9904  & 307029 \\
CAR110  & 10\uph42\upm15\ups & $-61$\arcd34\arcm35\arcs & 288{\zdot\arcd}1846 &  $-2${\zdot\arcd}4606  & 371079 \\
CAR111  & 10\uph47\upm15\ups & $-60$\arcd48\arcm45\arcs & 288{\zdot\arcd}3599 &  $-1${\zdot\arcd}5037  & 334938 \\
CAR112  & 10\uph52\upm20\ups & $-61$\arcd04\arcm15\arcs & 289{\zdot\arcd}0276 &  $-1${\zdot\arcd}4562  & 370525 \\
CAR113  & 10\uph57\upm20\ups & $-61$\arcd04\arcm08\arcs & 289{\zdot\arcd}5717 &  $-1${\zdot\arcd}1922  & 369094 \\
CAR114  & 10\uph57\upm20\ups & $-60$\arcd28\arcm18\arcs & 289{\zdot\arcd}3178 &  $-0${\zdot\arcd}6516  & 344169 \\
CAR115  & 10\uph40\upm30\ups & $-62$\arcd09\arcm00\arcs & 288{\zdot\arcd}2783 &  $-3${\zdot\arcd}0629  & 349173 \\
CAR116  & 10\uph37\upm00\ups & $-62$\arcd45\arcm00\arcs & 288{\zdot\arcd}2176 &  $-3${\zdot\arcd}7841  & 299996 \\
CAR117  & 10\uph42\upm05\ups & $-62$\arcd45\arcm00\arcs & 288{\zdot\arcd}7274 &  $-3${\zdot\arcd}5017  & 316077 \\
CAR118  & 10\uph38\upm30\ups & $-63$\arcd20\arcm50\arcs & 288{\zdot\arcd}6602 &  $-4${\zdot\arcd}2207  & 261809 \\
CEN106  & 11\uph32\upm30\ups & $-60$\arcd50\arcm00\arcs & 293{\zdot\arcd}4598 &  $ 0${\zdot\arcd}5784  & 500453 \\
CEN107  & 11\uph54\upm00\ups & $-62$\arcd00\arcm00\arcs & 296{\zdot\arcd}2458 &  $ 0${\zdot\arcd}1238  & 557251 \\
CEN108  & 13\uph33\upm00\ups & $-64$\arcd15\arcm00\arcs & 307{\zdot\arcd}4281 &  $-1${\zdot\arcd}7417  & 845638 \\
MUS100  & 13\uph15\upm00\ups & $-64$\arcd51\arcm00\arcs & 305{\zdot\arcd}4335 &  $-2${\zdot\arcd}0928  & 735652 \\
MUS101  & 13\uph25\upm00\ups & $-64$\arcd58\arcm00\arcs & 306{\zdot\arcd}4749 &  $-2${\zdot\arcd}3261  & 766910 \\
\noalign{\vskip3pt}
\hline}

\Section{Photometric Maps of the Galactic Disk Fields}
The construction of the OGLE-III maps was presented in detail in Udalski
\etal (2008a). We followed this procedure for the Galactic disk fields
as well.

The OGLE-III maps contain the mean photometry of all detected stellar
objects. The mean photometry was obtained for all objects with minimum of 6
observations in the {\it I}-band by averaging all observations after
removing $5\sigma$ outliers. Because of small number of {\it V}-band epochs
for some of the fields even a single {\it V}-band observation entered the
database. In the case of more {\it V}-band data points they were also
averaged with the $5\sigma$ outliers rejection.

\begin{landscape}
\renewcommand{\arraystretch}{0.95}
\MakeTableSepp{
r@{\hspace{15pt}}
c@{\hspace{15pt}}
c@{\hspace{15pt}}
r@{\hspace{15pt}}
c@{\hspace{15pt}}
c@{\hspace{15pt}}
r@{\hspace{15pt}}
c@{\hspace{15pt}}
c@{\hspace{15pt}}
r@{\hspace{15pt}}
c@{\hspace{15pt}}
r@{\hspace{15pt}}
c@{\hspace{15pt}}
c}{12.5cm}
{OGLE-III Photometric Map of the CAR100.2 field (sample)}
{\hline
\noalign{\vskip4pt}
ID & RA    & DEC   & $X$~~~~ & $Y$ & $V$ & $V-I$ & $I$ & $N_V$ & $N^{\rm bad}_V$ & $\sigma_V$ & $N_I$ & $N^{\rm bad}_I$ & $\sigma_I$\\
   &(2000) & (2000)&&&&&&&&&&&\\
\noalign{\vskip4pt}
\hline
\noalign{\vskip4pt}
     1 & 11\uph07\upm04\zdot\ups50 & $-61\arcd13\arcm44\zdot\arcs5$ & 456.05 &  53.37 & 15.624 & 1.631 & 13.993 & 3 & 0 & 0.001 & 1749 &  0 & 0.008\\
     2 & 11\uph07\upm06\zdot\ups55 & $-61\arcd13\arcm43\zdot\arcs3$ & 460.44 & 110.06 & 14.076 & 0.507 & 13.568 & 3 & 0 & 0.001 & 2533 & 10 & 0.009\\
     3 & 11\uph07\upm07\zdot\ups11 & $-61\arcd15\arcm00\zdot\arcs8$ & 162.74 & 123.92 & 18.553 & 4.137 & 14.416 & 3 & 0 & 0.022 & 2547 &  1 & 0.015\\
     4 & 11\uph07\upm08\zdot\ups69 & $-61\arcd13\arcm25\zdot\arcs4$ & 528.71 & 169.98 & 14.647 & 0.842 & 13.805 & 3 & 0 & 0.003 & 2667 & 12 & 0.011\\
     5 & 11\uph07\upm10\zdot\ups78 & $-61\arcd13\arcm26\zdot\arcs0$ & 525.92 & 227.89 & 14.392 & 9.999 & 99.999 & 3 & 0 & 0.003 &    0 &  0 & 9.999\\
     6 & 11\uph07\upm12\zdot\ups68 & $-61\arcd14\arcm47\zdot\arcs5$ & 213.08 & 278.48 & 16.273 & 2.599 & 13.674 & 3 & 0 & 0.007 & 2684 &  0 & 0.010\\
     7 & 11\uph07\upm13\zdot\ups88 & $-61\arcd14\arcm17\zdot\arcs0$ & 329.61 & 312.49 & 14.197 & 0.420 & 13.777 & 3 & 0 & 0.003 & 2682 &  2 & 0.006\\
     8 & 11\uph07\upm15\zdot\ups27 & $-61\arcd15\arcm09\zdot\arcs3$ & 129.01 & 349.62 & 15.406 & 1.741 & 13.665 & 3 & 0 & 0.002 & 2514 &  2 & 0.005\\
     9 & 11\uph07\upm16\zdot\ups33 & $-61\arcd15\arcm27\zdot\arcs8$ &  57.57 & 378.53 & 14.735 & 9.999 & 99.999 & 3 & 0 & 0.004 &    0 &  0 & 9.999\\
    10 & 11\uph07\upm06\zdot\ups24 & $-61\arcd15\arcm01\zdot\arcs8$ & 159.36 &  99.77 & 15.930 & 1.116 & 14.814 & 3 & 0 & 0.004 & 2371 &  3 & 0.007\\
    11 & 11\uph07\upm06\zdot\ups73 & $-61\arcd13\arcm47\zdot\arcs6$ & 443.65 & 115.12 & 16.710 & 1.101 & 15.609 & 3 & 0 & 0.009 & 2561 &  6 & 0.007\\
    12 & 11\uph07\upm10\zdot\ups87 & $-61\arcd14\arcm07\zdot\arcs8$ & 365.50 & 229.39 & 16.291 & 0.826 & 15.465 & 3 & 0 & 0.008 & 2681 &  3 & 0.006\\
    13 & 11\uph07\upm11\zdot\ups11 & $-61\arcd13\arcm57\zdot\arcs4$ & 405.35 & 236.04 & 16.061 & 0.980 & 15.081 & 3 & 0 & 0.005 & 2684 &  0 & 0.005\\
    14 & 11\uph07\upm11\zdot\ups69 & $-61\arcd13\arcm56\zdot\arcs7$ & 408.00 & 252.21 & 16.296 & 1.001 & 15.296 & 3 & 0 & 0.004 & 2674 & 10 & 0.009\\
    15 & 11\uph07\upm14\zdot\ups18 & $-61\arcd15\arcm20\zdot\arcs2$ &  87.43 & 318.99 & 17.938 & 2.442 & 15.496 & 3 & 0 & 0.002 & 2010 &  0 & 0.177\\
    16 & 11\uph07\upm15\zdot\ups33 & $-61\arcd15\arcm02\zdot\arcs8$ & 153.89 & 351.27 & 16.080 & 0.960 & 15.120 & 3 & 0 & 0.003 & 2603 &  0 & 0.006\\
    17 & 11\uph07\upm15\zdot\ups37 & $-61\arcd13\arcm27\zdot\arcs2$ & 520.71 & 354.97 & 15.932 & 0.926 & 15.006 & 3 & 0 & 0.005 & 2683 &  1 & 0.007\\
    18 & 11\uph07\upm17\zdot\ups08 & $-61\arcd14\arcm20\zdot\arcs9$ & 314.19 & 400.92 & 17.147 & 1.461 & 15.686 & 3 & 0 & 0.002 & 2683 &  1 & 0.007\\
    19 & 11\uph07\upm19\zdot\ups14 & $-61\arcd15\arcm13\zdot\arcs8$ & 111.06 & 456.46 & 16.629 & 1.004 & 15.625 & 3 & 0 & 0.004 & 2388 &  4 & 0.008\\
    20 & 11\uph07\upm20\zdot\ups06 & $-61\arcd13\arcm53\zdot\arcs3$ & 419.50 & 484.11 & 16.996 & 1.524 & 15.472 & 3 & 0 & 0.006 & 2682 &  2 & 0.006\\
    21 & 11\uph07\upm20\zdot\ups39 & $-61\arcd13\arcm36\zdot\arcs2$ & 485.25 & 493.84 & 16.486 & 0.898 & 15.587 & 3 & 0 & 0.002 & 2683 &  0 & 0.007\\
    22 & 11\uph07\upm20\zdot\ups54 & $-61\arcd14\arcm57\zdot\arcs2$ & 174.38 & 495.87 & 16.584 & 1.164 & 15.420 & 3 & 0 & 0.005 & 2653 &  1 & 0.008\\
    23 & 11\uph07\upm21\zdot\ups10 & $-61\arcd14\arcm19\zdot\arcs4$ & 319.17 & 512.36 & 16.438 & 0.844 & 15.593 & 3 & 0 & 0.005 & 2681 &  3 & 0.008\\
    24 & 11\uph07\upm04\zdot\ups20 & $-61\arcd15\arcm24\zdot\arcs0$ &  74.40 &  42.86 & 16.865 & 9.999 & 99.999 & 3 & 0 & 0.007 &    0 &  0 & 9.999\\
    25 & 11\uph07\upm04\zdot\ups22 & $-61\arcd14\arcm05\zdot\arcs1$ & 377.10 &  45.19 & 17.009 & 1.081 & 15.928 & 3 & 0 & 0.009 & 1589 &  0 & 0.009\\
\noalign{\vskip4pt}
\hline}
\end{landscape}

In Table~2 we present a sample of the data -- the first 25 entries from the
map of the CAR100.2 subfield. The columns contain: (1)~ID number;
(2,3)~equatorial coordinates J2000.0; (4,5)~$X,Y$ pixel coordinates in the
{\it I}-band reference image; (6,7,8)~photometry: {\it V}, $V-I$, {\it I};
(9,10,11)~number of points for average magnitude, number of $5\sigma$
removed points, $\sigma$ of magnitude for {\it V}-band; (12,13,14)~same as
(9,10,11) for the {\it I}-band. 9.999 or 99.999 markers mean ``no
data''. Value of $-1$ in column (9) indicates multiple {\it V}-band
cross-identification (when the {\it V}-band counterpart was detected in
more than one field in overlaping areas; the average magnitude is the mean
of the merged datasets).

The full set of the OGLE-III Photometric Maps of the Galactic Disk Fields
is available from the OGLE Internet archive (see Section~5).

\Section{Discussion}
The OGLE-III Photometric Maps of the Galactic Disk Fields contain entries
for about 9 million stars located in 21 OGLE-III fields. Figs.~1 and 2 show
the typical accuracy of the OGLE-III Photometric Maps of these targets:
standard deviation of magnitudes as a function of magnitude in the {\it V}-
and {\it I}-band for the field CAR100.2 and denser MUS100.2.

The completeness of the photometry can be assessed from the histograms
presented in Fig.~3 and 4 for the same fields as in Figs.~1 and 2. It
reaches $I\approx20.5$~mag and $V\approx 21$~mag.

Figs.~5--8 present color--magnitude diagrams (CMDs) constructed for a few
selected subfields from different Galactic disk fields observed by OGLE-III
survey.

\Section{Data Availability}
The OGLE-III Photometric Maps of the SMC are available to the
astronomical community from the OGLE Internet Archive:

\begin{center}
{\it http://ogle.astrouw.edu.pl}\\
{\it ftp://ftp.astrouw.edu.pl/ogle3/maps/gd/}
\end{center}

The archives include, besides the tables with photometric data and
astrometry for each of the subfields, also the {\it I}-band reference
images. Usage of the data is allowed under the proper acknowledgment to
the OGLE project.

We also plan to build an on-line, interactive access to the photometric
maps database, allowing to search for objects fulfilling user-defined
set of criteria. The availability of such an access will be announced on
the OGLE project WWW page.

\Acknow{The OGLE project has received funding from the European Research
Council under the European Community's Seventh Framework Programme
(FP7/2007-2013) / ERC grant agreement no. 246678 to AU.}

\vspace{8mm}
\noindent List of figures.
\vspace{2mm}

\FigCap{Standard deviation of magnitudes as a function of magnitude for
the field CAR100.2.}
\FigCap{Standard deviation of magnitudes as a function of magnitude for
the field MUS100.2.}
\FigCap{Histogram of stellar magnitudes in the Galactic disk field CAR100.2.}
\FigCap{Histogram of stellar magnitudes in the Galactic disk field MUS100.2.}
\FigCap{Color--magnitude diagram of the Galactic disk field CAR100.2.}
\FigCap{Color--magnitude diagram of the Galactic disk field CAR110.2.}
\FigCap{Color--magnitude diagram of the Galactic disk field CEN108.2.}
\FigCap{Color--magnitude diagram of the Galactic disk field MUS100.2.}

\end{document}